\documentclass{camera}
\usepackage{graphicx} 

\begin{document}

%
\title{Open Charm Mesons at the LHC with ALICE}
%
\author{Renu Bala \\ for the ALICE Collaboration}


\organization{Universit\`a  di Torino e INFN Torino}
\maketitle
\begin{abstract}
 The ALICE experiment will be able to detect hadrons containing charm and beauty quarks in proton-proton and heavy ion collisions in the new energy regime of the CERN Large Hadron Collider (LHC). Open charmed mesons are a powerful tool to study the medium produced in heavy ion collisions, since charm quarks are produced on a very short time scale and they experience the whole history of the collision. In addition, the measurements of heavy flavour yield provide a natural normalization for those of charmonia and bottomonia production at LHC. In this talk, after a general overview of ALICE perspectives for heavy flavour physics, we will report some study of D-meson reconstruction through their hadronic decay channels with Monte Carlo simulated data.
\end{abstract}

%
\section{Introduction}
The nucleon-nucleon centre of mass energy for the Pb-Pb collisions at LHC, $\sqrt{s_{NN}}$ = 5.5 TeV, will exceed that available at RHIC (the Relativistic Heavy Ion Collider) by a factor about 30, opening up a new domain for the study of strongly interacting matter in conditions of high temperature (factor 3 to 5 higher than the critical temperature) and high energy density (15 to 60 GeV/$fm^3$). QCD calculations[1] at high temperature and high energy density predict a phase transition from hadron gas to Quark-Gluon Plasma (a deconfined state of matter). In the study of the properties of the produced (deconfined) state, heavy quarks play a crucial role. Heavy quarks and hard partons, abundantly produced at LHC energies in the initial hard scattering processes, are sensitive probes of the medium formed in the collisions as they may lose energy by gluon radiation while propagating through the medium itself.  At LHC energies ($\sqrt{s}$ = 14 TeV for p-p and $\sqrt{s_{NN}}$ = 5.5 TeV for Pb-Pb collisions), the production of charm and beauty will be abundant. The cross-sections at LHC are expected to  increase by about a factor of 10 for charm and 100 for beauty with respect to RHIC. The baseline production cross-section of Q$\bar{Q}$ pairs for ALICE simulation studies has been calculated in the framework of collinear factorization and pQCD[2], including the nuclear modification of the parton distribution functions (PDFs)[3]. The expected c$\bar{c}$ and b$\bar{b}$ production yields for pp collisions at $\sqrt{s}$ = 14 TeV are 0.16 and 0.0072, respectively[4]. For the 5 \% most central Pb-Pb collisions at $\sqrt{s_{NN}}$ = 5.5 TeV, the expected yields are 115 and 4.6 respectively. It has to be noted that these predictions have large uncertainties, of about a factor 2 to 3, depending on the choice of the quark masses and QCD scales.
\section{ALICE at the LHC}
 The ALICE experimental setup, described in detail in [4], has excellent capabilities for heavy flavour measurements. Experimentally, the two key elements for a rich heavy-flavour program are: tracking/vertexing and particle identification. {\bf Tracking/vertexing}: Particle tracking relies on the six concentric layers of high resolution silicon detectors of the Inner Tracking System (ITS)\footnote{Two innermost layers equipped with silicon pixel (SPD) plus two layers of silicon drift detectors (SDD), and two layers of silicon strip detectors (SSD)}, a large volume of time projection chamber (TPC)\footnote{A large drift chamber with multiwire proportional chamber (MWPC) endcaps[4]}, and a high granularity transition-radiation detector (TRD). The ALICE detection strategy for charm and beauty hadrons relies on resolving secondary detached vertices reconstructed from tracks with large impact parameters ($d_0$)\footnote{the impact parameter being the distance of closest approach of a particle trajectory to the primary vertex}. The precision in impact parameter measurement is mainly provided by: two innermost ITS layers (SPD) in the bending plane (r$\phi$) \& two intermediate ITS layer (SDD) for z-coordinate. A resolution ($\sigma_{d_0}$) better than 60$\mu$m in the bending plane is achieved for tracks with $p_t > $ 1.5 GeV/c. {\bf Particle Identification}: In the central region, particle identification is performed over the full azimuth by a dE/dx measurement in the tracking detectors (TPC and ITS), via time of flight measurement using Time of Flight detector(TOF) and transition radiation measurement in the Transition Radiation Detector (TRD).
\section{Charm Reconstruction in the hadronic decays}
  An intensive simulation study of the ALICE performance for the reconstruction of D mesons from hadronic decays has been already done using the decay channels $D^0 \rightarrow K^- \pi^+$ [5,6], $D^+ \rightarrow K^- \pi^+ \pi^+ $[7] and $D_s \rightarrow K^- K^+ \pi^+ $ whereas  other channels like $D^* \rightarrow D^0 \pi$, $D^0 \rightarrow K \pi \pi \pi $ and $\Lambda_c \rightarrow \pi$ Kp are under study. It is important to measure the production of as many as charm hadrons as possible because the measurement of their relative yield (e.g $D^{+}_{s}$(c$\bar{s}$)/$D^+$(c$d^-$)) can provide information on the hadronization mechanism (e.g string fragmentation and recombination) and it allows to reduce the systematic error on the absolute cross-section. 

{\bf Charm Reconstruction in the $D^0 \rightarrow K^- \pi^+$ decay channel:}
  The first study in ALICE on open charmed mesons was performed on $D^0$ mesons through their two body hadronic channel $K^- \pi^+$, which is the simplest decay topology. Two main variables are used to separate the signal from the combinatorial background of opposite sign track pairs: the product of the impact parameters of the two tracks ($d_{0}^{K} \times d_{0}^{\pi}$) and the cosine of the pointing angle ($\theta_{point}$)\footnote{The angular distance between the reconstructed $D^0$ momentum and the $D^0$ flight line}. By performing a selection based on these two variables, it is possible to improve the signal to background ratio by more than a factor of $10^3$.
\begin{figure}[h!]
 \centering
  \includegraphics[width=0.45\textwidth]{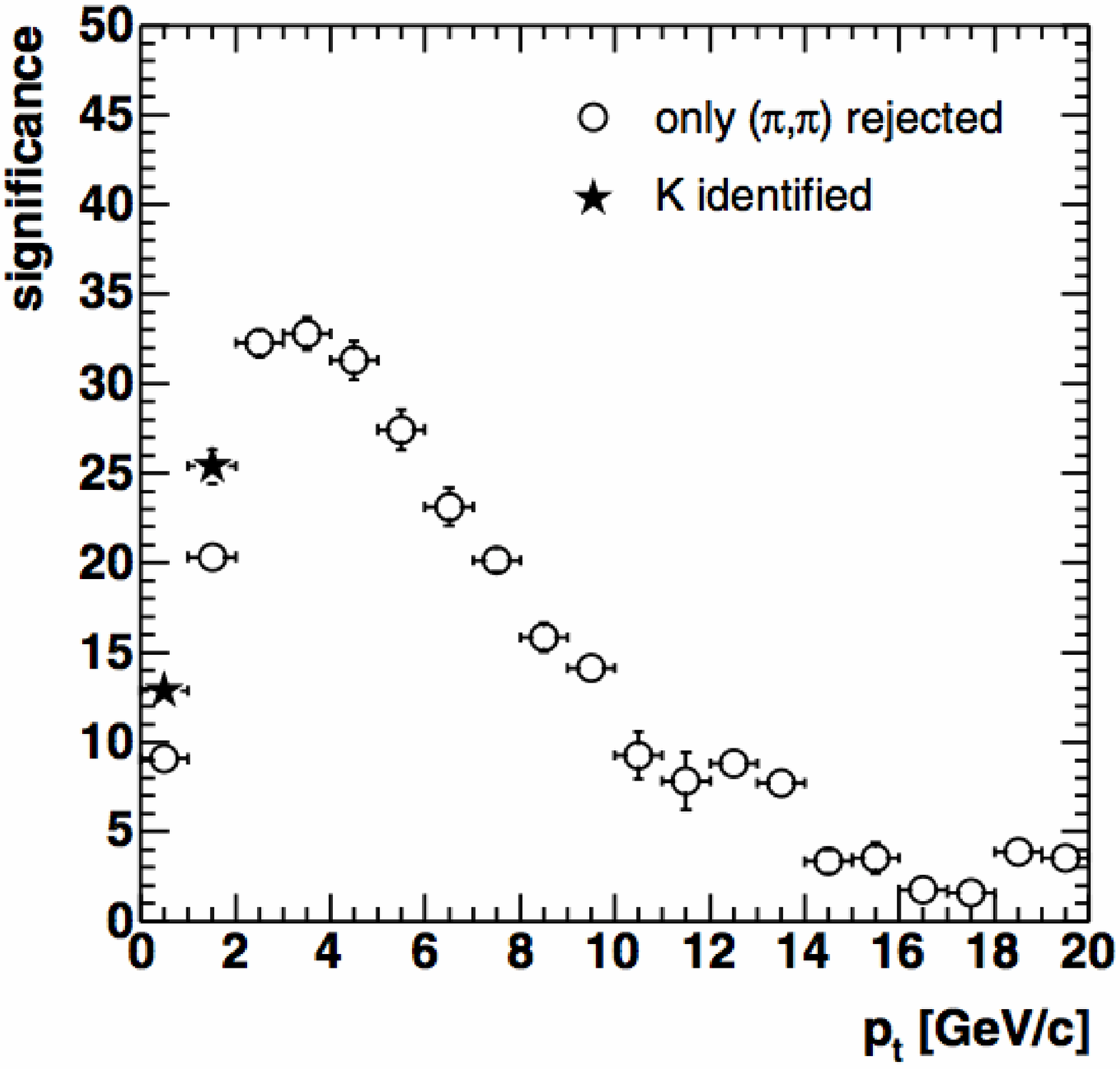}
  \includegraphics[width=0.45\textwidth]{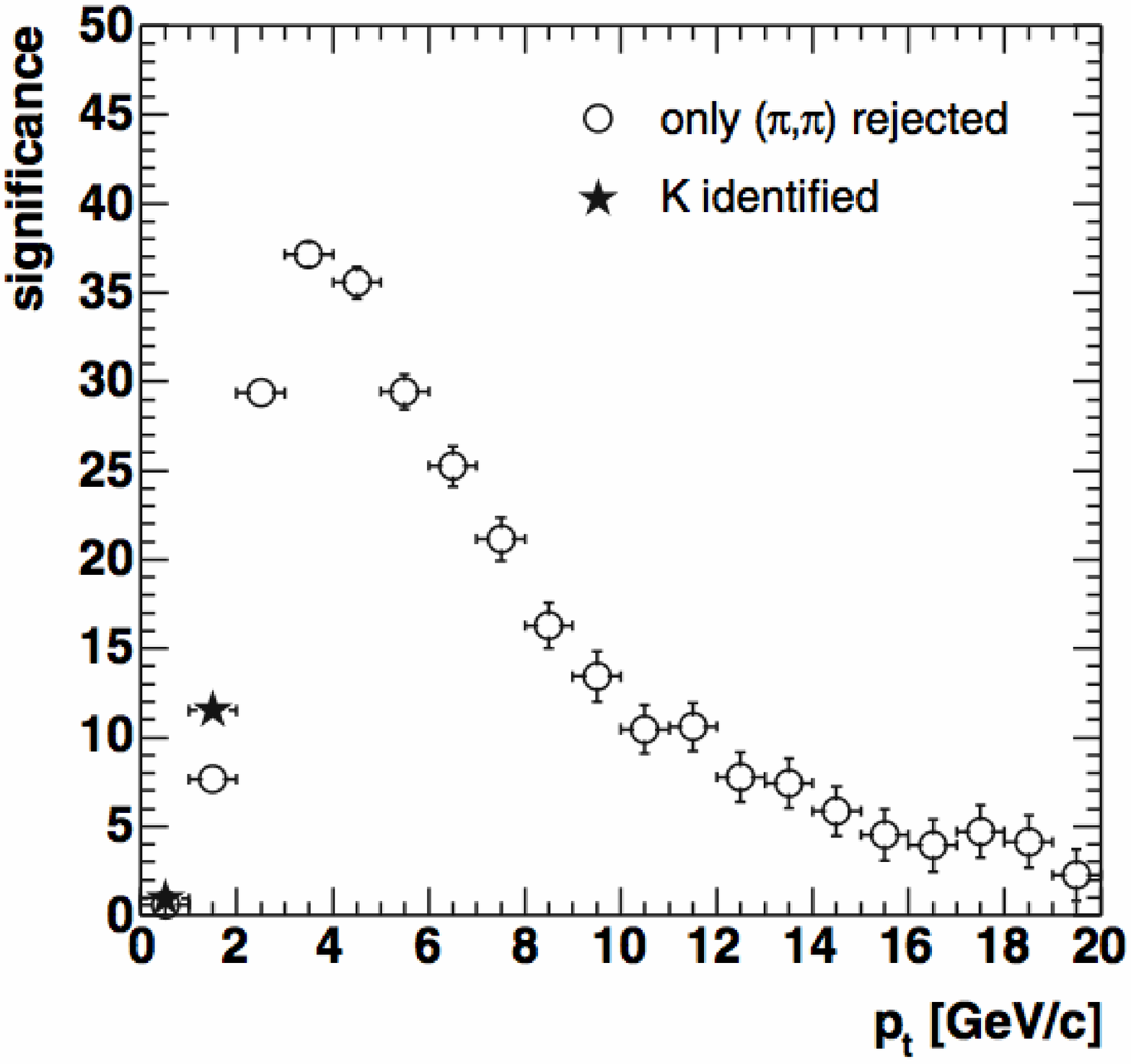}
  \caption{\it Significance as a function of $D^0$ transverse momentum for Pb-Pb (right) and pp (left) collision[5,6].}
  \label{d0}
\end{figure}
 Figure~\ref{d0} shows the significance(S/$\sqrt{S+B}$) as a function of $p_t$ in bins of 1 GeV/c. The relative statistical error on total $p_{t}$ integrated charm cross-section in one year of data taking at nominal luminosity are expected to be smaller than 20\% [8].\\
{\bf Charm Reconstruction in the $D^+ \rightarrow K^- \pi^+ \pi^+$ decay channel: }
    Extracting a $D^+$ signal with a good significance requires a drastic selection procedure to reduce the huge combinatorial background by at least 8-9 orders of magnitude. The main variables to separate the signal from huge combinatorial background of the three charged track combinations are: the distance between the primary and secondary vertex and the cosine of pointing angle ($\theta_{point}$).
\begin{figure}[hptb]
\centering
   \includegraphics[width=1.0\textwidth]{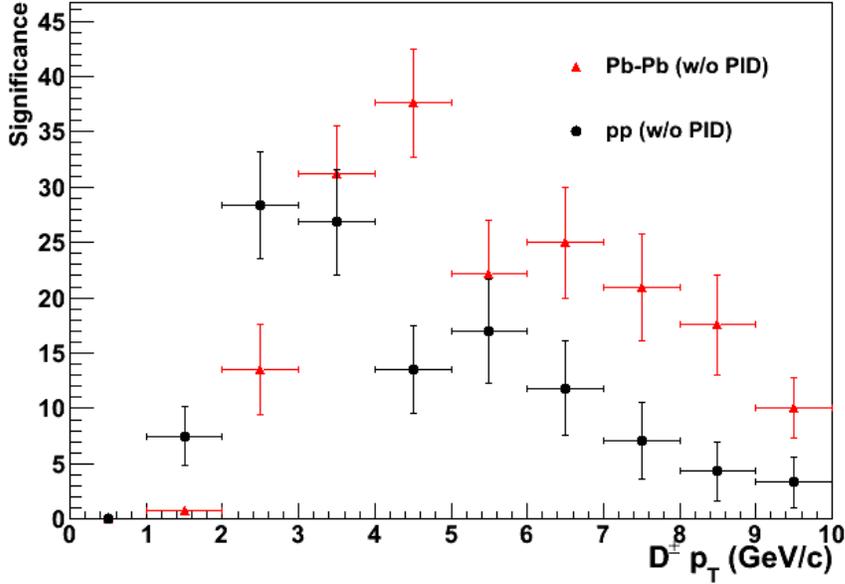}
  \caption{\it Significance as a function of $D^\pm$ transverse momentum for pp ($10^9$ events) and Pb-Pb ($10^7$ events)}
  \label{significance}
\end{figure}
If the found vertex really corresponds to the a $D^+$ vertex, then $\theta_{point} \approx$ 0 and Cos$\theta_{point} \approx$ 1.
 The significance as a function of $D^{\pm}$ transverse momentum in bins of 1 GeV/c is shown in figure~\ref{significance}. The plot confirms that there is good feasibility of reconstruction of $D^\pm$ mesons in pp and Pb-Pb collision in the first year of data taking even in the worst scenario where the particle identification information is not used.\\
{\bf Charm Reconstruction in the $D_s \rightarrow K^- K^+ \pi^+$ decay channel: }
   The reconstruction of $D_s$ meson is a very challenging task as the yield of $D_s$ against the huge combinatorial background is very low. Nevertheless, $D_s$ mesons preferentially decay through intermediate resonant states and this fact can improve the separation of signal from background. From the simulation studies of this decay channel (see [9]), we observed that the reconstruction of $D_s$ mesons is feasible for $p_{t}$ down to 3-4 GeV/c, even without particle identification, by exploiting  the $D_{s}^{+} \rightarrow \phi \pi^+ \rightarrow K^- K^+ \pi^+$ decay.
\section{Charm Energy Loss: Nuclear Modification factor}
  The measured spectra in pp and Pb-Pb can be used to compute the nuclear modification factor, $R_{AA}(p_t) = \frac{d^2 N_{AA}/dp_t dy}{<N_{coll}> d^2N_{pp}/dp_t dy}$. This observable is supposed to be 1 if the nucleus-nucleus collision behaves as a simple superposition of independent nucleon-nucleon collisions.
\begin{figure}[h!]
 \centering                                                                                                                                                    
  \includegraphics[width=1.0\textwidth]{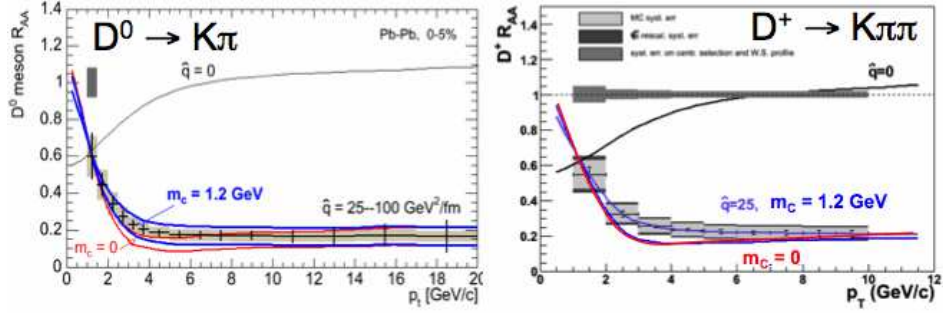}
  \caption{\it Expected performance for the measurement of Nuclear Modification factor of $D^0$ mesons (left) and $D^+$ (right) after one year of data taking at nominal luminosity }
  \label{energy_loss}
\end{figure}
The expected performance for the measurement of the nuclear modification factor for $D^0$ and $D^+$ mesons after one year of data taking  at nominal luminosity is shown in figure~\ref{energy_loss}. Theoretical calculations for different energy loss scenarios depending on the in-medium transport coefficient $\hat{q}$ and on the c-quark mass are also shown.  The bands corresponding to $m_c$ =1.2 GeV and $\hat{q}$= 25-100 $GeV^2$/fm reflects the estimated uncertainty on the model expectations for $R_{AA}^D$. The small difference between the two bands ($m_c$=0 and $m_c$=1.2 GeV) indicates that with respect to energy loss, charm behaves similarly to light quarks. Therefore, the enhancement of the heavy to light ratio is a sensitive measurement, essentially free of mass effects, to study the colour charge dependence of parton energy loss[10].
\section{Charm Flow}
   Flow, in particular elliptic flow[11], is an important experimental probe which provides  information about thermalization of the medium created in  non-central heavy ion collision. Here we will show the expected performance for the measurement of  $D^+$ $v_2$ as a function of $p_t$ with ALICE.
    \begin{figure}[h!]
   \centering                                                                                                                                                  
  \includegraphics[width=1.0\textwidth]{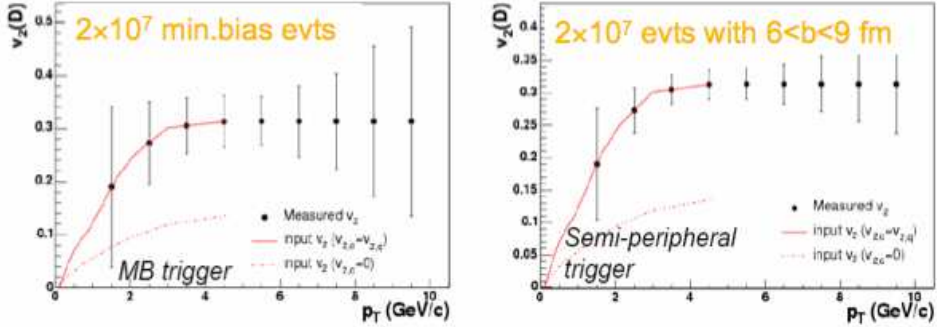}

  \caption{\it $D^\pm$ meson $v_2$ as a function of $p_t$ in centrality class $ 6<b<9$ fm.}
  \label{flow}
\end{figure}
As shown in figure~\ref{flow}, we have compared two cases: Minimum bias trigger (left panel) and Semi-peripheral trigger (right panel). The lines represent the prediction for $D^\pm$ meson $v_{2}^{D}(p_t)$ in the case of charm flow and without charm flow. The statistical errors on $v_{2}^D$ in 2 $\times 10^7$ minimum bias events are pretty large whereas in case 2$\times 10^7$ semi-peripheral events with 6 $< b < $ 9 fm, the statistical errors appreciably reduced and thus, $v_2$ as a function of $p_t$ is significally improved down to $p_t \sim $ 1 GeV/c.
\section{Conclusions}
 With its excellent tracking, vertexing and particle identification capabilities, ALICE has very promising prospectives for the studies of charm production in the hadronic decays. We have shown that, there is good feasibility of reconstruction of D-mesons with a good significance both in Pb-Pb and pp collisions in one year of data taking at nominal luminosity. This will allow to measure with good precision the energy loss and the azimuthal anisotropy of the charm particles.

%

\end{document}